\documentclass{aa}
\usepackage{graphicx}
\usepackage{txfonts}
\usepackage{natbib}
\bibpunct{(}{)}{;}{a}{}{,} 

\begin{document}
\newcommand{\kep}{{{\it Kepler }}}
\newcommand{\kic}{{{KIC~8429280 }}}

\newcommand{\block}{\vrule width 0.5 true cm height 6pt depth 0pt \ }

\title{Magnetic activity and differential rotation in the very young star KIC~8429280\thanks{Based on public \kep data, on 
observations made with the Italian Telescopio Nazionale Galileo operated on the island of La Palma by the Centro 
Galileo Galilei of INAF (Istituto Nazionale di Astrofisica) at the Spanish Observatorio
del Roque del los Muchachos of the Instituto de Astrofisica de Canarias, and on observations collected at the
Catania Astrophysical Observatory (Italy).}}
\author{A. Frasca\inst{1}\and
	H.-E. Fr\"ohlich\inst{2}\and
	A. Bonanno \inst{1}\and
	G. Catanzaro\inst{1}\and
	K. Biazzo\inst{3,1}\and
	J. Molenda-\.Zakowicz\inst{4}
          }

\offprints{A. Frasca\\ \email{antonio.frasca@oact.inaf.it}}

\institute{
INAF, Osservatorio Astrofisico di Catania, via S. Sofia, 78, 95123 Catania, Italy
\and
Leibniz Institute for Astrophysics Potsdam (AIP), An der Sternwarte 16, 14482 Potsdam, Germany
\and
INAF, Osservatorio Astronomico di Capodimonte, via Moiariello, 16,  I--80131 Napoli, Italy
\and 
Astronomical Institute, Wroc{\l}aw University, ul.\ Kopernika 11, 51-622 Wroc{\l}aw, Poland
}

\date{Received 30 March 2011 / Accepted 03 June 2011}

 
\abstract 
{}
{We present a spectroscopic and photometric analysis of the rapid rotator KIC~8429280, discovered by ourselves as a very young 
star and observed by the NASA \kep mission, designed to determine its activity level, spot distribution, and 
differential rotation.}
{We use ground-based data, such as high-resolution spectroscopy and multicolor broad-band photometry, to derive	
stellar parameters ($v\sin i$, spectral type, $T_{\rm eff}$, $\log g$, and [Fe/H]), and we adopt a spectral subtraction technique 
to highlight the strong chromospheric emission in the cores of hydrogen H$\alpha$ and \ion{Ca}{ii} H\&K and infrared triplet (IRT) lines. 
We then fit a robust spot model to the high-precision \kep photometry spanning 138 days.
Model selection and parameter estimation is performed in a Bayesian manner using 
a Markov chain Monte Carlo method.  }
{We find that KIC~8429280 is a cool (K2\,V) star with an age of about 50\,Myr, based on its lithium content, that
has passed its T Tau phase and is spinning up approaching the ZAMS on its radiative track.  Its high level of 
chromospheric activity is clearly indicated by the strong radiative losses in \ion{Ca}{ii} H\&K and IRT,
H$\alpha$, and H$\beta$ lines. Furthermore, its Balmer decrement and the flux ratio of \ion{Ca}{ii} IRT  lines imply
that these lines are mainly formed in optically-thick sources analogue to solar plages.
The analysis of the \kep data uncovers evidence of at least 
seven enduring spots. 
Since the star's inclination is rather high -- nearly $70^\circ$ -- the assignment of the spots to either the northern 
or southern hemisphere is not unambiguous. We find at least three solutions with nearly the same 
level of residuals. 
Even in the case of seven spots, the fit is far from being perfect. Owing to the exceptional precision of the \kep photometry, 
it is not possible to reach the noise floor without strongly enhancing the degrees of freedom and,
consequently, the non-uniqueness of the solution.
The distribution of the active regions is such that the spots  are located around three latitude 
belts, i.e. around the star's equator and around $\pm (50{\degr}$--60${\degr})$, with the high-latitude spots rotating slower 
than the low-latitude ones.
The equator-to-pole differential rotation $d\Omega\simeq 0.27$ rad\,d$^{-1}$ 
is at variance with some recent mean-field models of differential rotation in rapidly rotating main-sequence stars, which predict a much 
smaller latitudinal shear. Our results are consistent with the scenario of a higher differential rotation, which changes along the magnetic cycle, 
as proposed by other models. }
{}
\titlerunning{Magnetic activity and differential rotation in KIC~8429280}
      \authorrunning{A. Frasca et al.}

\maketitle

\section{Introduction}
\label{Sec:intro}

Starspots are characteristics of solar-like activity observed in cool stars.
They are tracers of magnetic flux tube emergence and can provide information about the different forces 
acting on the flux tubes during their buoyant rise, and on photospheric motions, such as the latitudinal 
drift of spots during the activity cycle and the differential rotation, which are both essential ingredients of mean
field dynamo theory. 

However, even though the activity cycle in Sun-like stars is believed to be driven by the meridional circulation with a low
eddy diffusivity \citep{Bonanno02}, the situation is much less clear for late-type rapidly rotating stars.
In this latter case, numerical simulations suggest that  
the magnetic dynamo action can produce wreaths of strong toroidal magnetic field at low latitudes, 
often of opposite polarity in the two hemispheres \citep{Nelson11}.
On the other hand, theoretical models of generation of magnetic fields in fully convective pre-main sequence (PMS) stars 
argue that the basic dynamo action is in the $\alpha^2$ regime \citep{Kueker99}. 

The differential rotation pattern is expected to be dictated by the Taylor-Proudman theorem, so that the iso-contour lines
are cylinder-shaped. 

Recent calculations of a 1-$M_{\sun}$ rapidly rotating main-sequence star with period $P=1.33$ days, predict 
a value of 0.08 rad\,d$^{-1}$, surprisingly close to the Solar value despite there being a factor of 20 between
the average rotation rates \citep{Kueker11}.
However, for \object{HD~171488}, a young Sun with an equatorial rotation period of 1.33\,days, values of differential 
rotation as high as 0.50 rad\,d$^{-1}$ have been found \citep[e.g., ][]{Marsden06,Jeffers08}. 

A unique opportunity for trying to solve these open problems
is given by the NASA \kep mission, which is providing an unprecedented data set 
of the photometric variability of a large star sample \citep{Borucki10}. Its combination of very high
photometric precision and  long, uninterrupted time coverage is essential not only for its primary goal (the discovery of 
transiting exoplanets) but also for the study of many astrophysical phenomena in stellar variability, such as the magnetic 
activity and differential rotation in cool stars.

Single active stars in the field can be most efficiently selected on the basis of their high coronal emission, and turn 
out to be mostly young stars with an age of a few hundred Myr, i.e. in the zero-age main sequence (ZAMS), or 
even younger, namely weak T Tauri (wTTS) or post-T Tauri (pTTS) stars \citep[see, e.g.,][and reference therein]{Guillout09}.

\object{KIC~8429280} (=\object{TYC~3146-35-1}) was selected as an X-ray active star from the cross-correlation of the ROSAT 
All-Sky Survey (RASS; \citealt{Voges1999, Voges2000}) with Tycho and Hipparcos catalogs \citep{HIPPA97}. 
It is a relatively bright star ($V=9.9$\,mag) and falls in the field of view of {\it Kepler}.	

\section{Ground-based observations and data reduction}
\label{Sec:Data}

\subsection{Spectroscopy}
Three spectra of KIC~8429280 were acquired with SARG, the {\it \'echelle} spectrograph at the Italian 
TNG telescope (La Palma, Spain). The first one was taken on 2007 May 31 within a survey of 
optical counterparts to X-ray sources (proposal TAC67-AOT15/07A).
With the yellow grism and a slit width of 0$\farcs$8, this spectrum has a resolution of $R=\lambda/\Delta\lambda\simeq 57\,000$, 
covering the spectral range 4600--7900\,\AA\  in 55 {\it \'echelle} orders, with a signal-to-noise ratio S/N$\sim$70.
The other two spectra (S/N ranging from about 60 to 100) were taken, adopting the same slit width, on 2009 Aug 11  
(proposal TAC71-AOT20/09B) with the red and blue grisms and cover the wavelength ranges 5500--11\,000\,\AA\  and 3600--5100\,\AA, 
respectively.

Another spectrum of KIC~8429280 with S/N$\sim$50 was taken at the $M. G. Fracastoro$ station (Serra La Nave, 
Mt. Etna, 1750 m a.s.l.) of the \textit{Osservatorio Astrofisico di Catania} (OAC - Italy) on 2009 July 13. 
The 91-cm telescope of the OAC was equipped with FRESCO, a fiber-fed {\it \'echelle} spectrograph that
covers the spectral range 4300--6800\,\AA\  in 20 orders with a resolution $R\simeq\,21\,000$. 

For calibration purposes, spectra of radial and rotational velocity standard stars (see Table\,\ref{Tab:Standards}), as well as bias, 
flat-field, and arc-lamp exposures were acquired at both observatories during each observing run. 

The data reduction was performed with the {\sc echelle} task of the IRAF\footnote{IRAF is distributed by the 
National Optical Astronomy Observatory, which is operated by the Association of the Universities for Research in 
Astronomy, inc. (AURA) under cooperative agreement with the National Science Foundation.} package, following 
standard steps \citep[see, e.g.,][]{Catanza10}.

\begin{table}[htb]
\caption{ Radial/rotational velocity standard stars.}
\begin{tabular}{llrcll}
\hline
\hline
Name       & Sp. Type &   $RV$             & $v\sin i^{\mathrm{d}}$      & Notes\\  
           &          &   (km\,s$^{-1}$)   & (km\,s$^{-1}$) &      \\  
\hline
\noalign{\smallskip}
HD\,115404 &  K2\,V   &   7.60$^{\mathrm{a}}$   & 3.3  & $RV$ \\  
HD\,145675 &  K0\,V   &    ...~~~~              & 0.8  & $v\sin i$ \\  
HD\,157214 &  G0\,V   &  $-79.2^{\mathrm{b}}~ $ & 1.6  & $v\sin i$ \\  
HD\,10700  &  G8\,V   &  $-17.1^{\mathrm{b}}~ $ & 0.9  & $v\sin i$ \\  
HD\,221354 &  K1\,V   &  $-25.20^{\mathrm{a}}$  & 0.6  & $RV$, $v\sin i$ \\  
HD\,32923  &  G4\,V   &   20.50$^{\mathrm{a}}$  & 1.5  & $RV$, $v\sin i$ \\  
HD\,12929  &  K2\,III & $-14.6^{\mathrm{c}}~ $  & 1.6  & $RV$ \\  
HD\,187691 &  F8\,V   &  $-0.2^{\mathrm{c}}~ $  & 2.8  & $RV$ \\  
HD\,182572 &  G8\,IV  & $-100.35^{\mathrm{a}}$  & 1.9  & $RV$, $v\sin i$ \\    
HD\,161096 &  K2\,III & $-12.53^{\mathrm{a}}$   & 2.1  & $RV$  \\
\hline
\end{tabular}
\label{Tab:Standards}
\begin{list}{}{}									
\item[$^{\mathrm{a}}$] \citet{Udry}. $^{\mathrm{b}}$ \citet{Nordstr}. 
\item[$^{\mathrm{c}}$] \citet{Evans1967}. $^{\mathrm{d}}$ \citet{Glebocki2005}.
\end{list}
\end{table}

\subsection{Photometry}

CCD images in the Johnson-Cousins $B$, $V$, $R_{\rm C}$, and $I_{\rm C}$ filters were acquired using the 
focal-reducer imaging camera with the 91-cm telescope of the OAC on 2010 November 11. 
Data reduction was carried out following standard steps of overscan region subtraction, master-bias subtraction, 
and division by average twilight flat-field images. 
The $BVR_{\rm C}I_{\rm C}$ magnitudes were extracted from the corrected images through aperture photometry 
performed with DAOPHOT using the IDL\footnote{IDL (Interactive Data Language) is a registered trademark 
of ITT Visual Information Solutions.} routine \textsc{APER}. 
Standard stars in the cluster \object{NGC~7790} \citep{Stetson2000} were observed to 
calculate the transformation coefficients to the Johnson-Cousins system. The zero points for the
$V$ magnitude and $B-V$, $V-R$, and $R-I$ colors were determined by the observation of a standard star close
to KIC\,8429280 (\object{BD+42\,3339}, \citealt{Castelaz}).
Photometric data are summarized in Table~\ref{Tab:StarPhot}.

%
\begin{table*}
\caption{Johnson-Cousins photometric data. Uncertainties are in parenthesis.}
\begin{tabular}{lccccccc}
\hline
\hline
Name          &  $B$          &  $V$          & $R_{\rm C}$   & $I_{\rm C}$    &   $J^{\rm a}$   &   $H^{\rm a}$    &   $K_{s}^{\rm a}$  \\
\hline
 KIC~8429280  &  10.88 (0.04) &  9.94  (0.04) &  9.38 (0.05)  &   8.86  (0.06) &   8.114 (0.021) &   7.672 (0.031)  &   7.541  (0.018) \\
\hline
\end{tabular}
\label{Tab:StarPhot}
\begin{list}{}{}		         	                	                 
\item[$^{\mathrm{a}}$]  From 2MASS catalog \citep{2MASS}.  
\end{list}
\end{table*}

\begin{figure}[htp]
\hspace{-0.3cm}
\includegraphics[width=9.5cm]{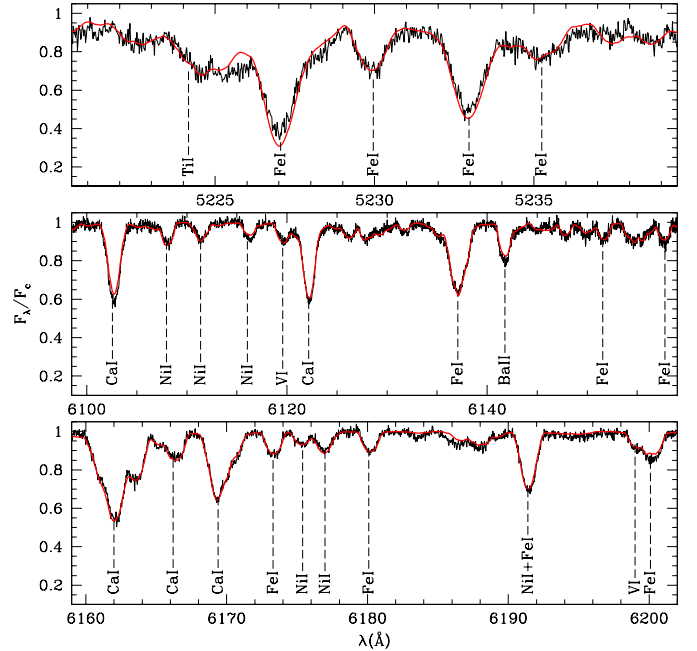}	
\vspace{-0.6cm}
\caption{Continuum-normalized observed spectrum (solid black line) in the region between 5220 and 5240 \AA ~({\it top panel})
and between 6100 and 6200 \AA ~({\it middle and bottom panels}) of KIC~8429280. 
The synthetic spectrum calculated with the parameters listed in Table~\ref{Tab:StarChar} is superimposed (smooth red line).}
\label{Fig:synthe}
\end{figure}

\section{Target characterization from ground-based observations}
\label{Sec:Char}

\subsection{Astrophysical parameters}
\label{Sec:Analysis}

High-resolution spectroscopy permits us to derive most of the fundamental stellar parameters, such 
as radial ($RV$) and projected rotational velocity ($v\sin i$), spectral type, luminosity class, effective 
temperature ($T_{\rm eff}$), gravity ($\log g$), and metallicity ([Fe/H]). 

Cross-correlation functions (CCFs) were computed with the IRAF task \textsc{fxcor} to derive both $RV$ and 
$v\sin i$. For this purpose, as templates we used spectra of slowly rotating $RV$ standard stars listed in 
Table\,\ref{Tab:Standards}.
We averaged the results from individual {\it \'echelle} orders as described, e.g., in \citet{Frasca2010}.
The CCFs were used for the determination of $v\sin i$ through a 
calibration of the full-width at half maximum (FWHM) of the CCF peak as a function of the $v\sin i$ of artificially 
broadened spectra of slowly rotating standard stars (Table\,\ref{Tab:Standards}) acquired with the same setup and 
in the same observing seasons as \kic  (see, e.g., \citealt{Guillout09}), obtaining a value of 39.3$\,\pm$\,2.4 km\,s$^{-1}$, 
where the error is the standard deviation of $v\sin i$ values from each individual {\it \'echelle} order.

We measured, within the errors, the same $RV$ in the SARG spectra of \kic taken in 2007 and 2009 ($RV=-33.1\pm 0.5$\,km\,s$^{-1}$,
and $-33.1\pm 0.4$\,km\,s$^{-1}$, respectively) and the FRESCO spectrum of 2009 ($RV=-32.2\pm 0.8$\,km\,s$^{-1}$).
The constant $RV$ measured by ourselves and the absence of any sign of duplicity in both the CCF and the line 
spectrum, suggest that KIC~8429280 could be a single star, although more spectra and high angular resolution 
images are needed to safely exclude the presence of a stellar companion (see Sect.~\ref{Sec:Disc:bin} for a wider discussion).

\begin{table*}[t]
\caption{Astrophysical parameters. Uncertainties are in parenthesis.}
\begin{tabular}{llccccccc}
\hline
\hline
Name      & Sp. Type   & $RV$	        &  $T_{\rm eff}$ & $\log g$    & [Fe/H]         & $v\sin i$      & $W_{\rm Li}$ & $\log N{\rm(Li)}$\\
          &            & (km\,s$^{-1}$) & (K)            &             &                & (km\,s$^{-1}$) & (m\AA)   &	 \\
\hline
\noalign{\smallskip}
KIC~8429280 & K2\,V  &  $-33.1$ (0.5)  &  5055 (135)$^{\rm a}$     & 4.41 (0.25)$^{\rm a}$ & $-0.02$ (0.10)$^{\rm a}$  & 37 (3)$^{\rm a}$  &  270 (20)   & 2.9 (0.1)  \\
            &        &                 &  5000 (100)$^{\rm b}$     & 4.50 (0.10)$^{\rm b}$ & $-0.05$ (0.10)$^{\rm b}$  & 39 (3)$^{\rm b}$  &		 &	   \\
\hline
\end{tabular}
\label{Tab:StarChar}
\begin{list}{}{}									
\item[$^{\mathrm{a}}$] From ROTFIT.  
\item[$^{\mathrm{b}}$] From SYNTHE.
\end{list}
\end{table*}

\begin{figure*}[th]
\centering
\includegraphics[width=15cm]{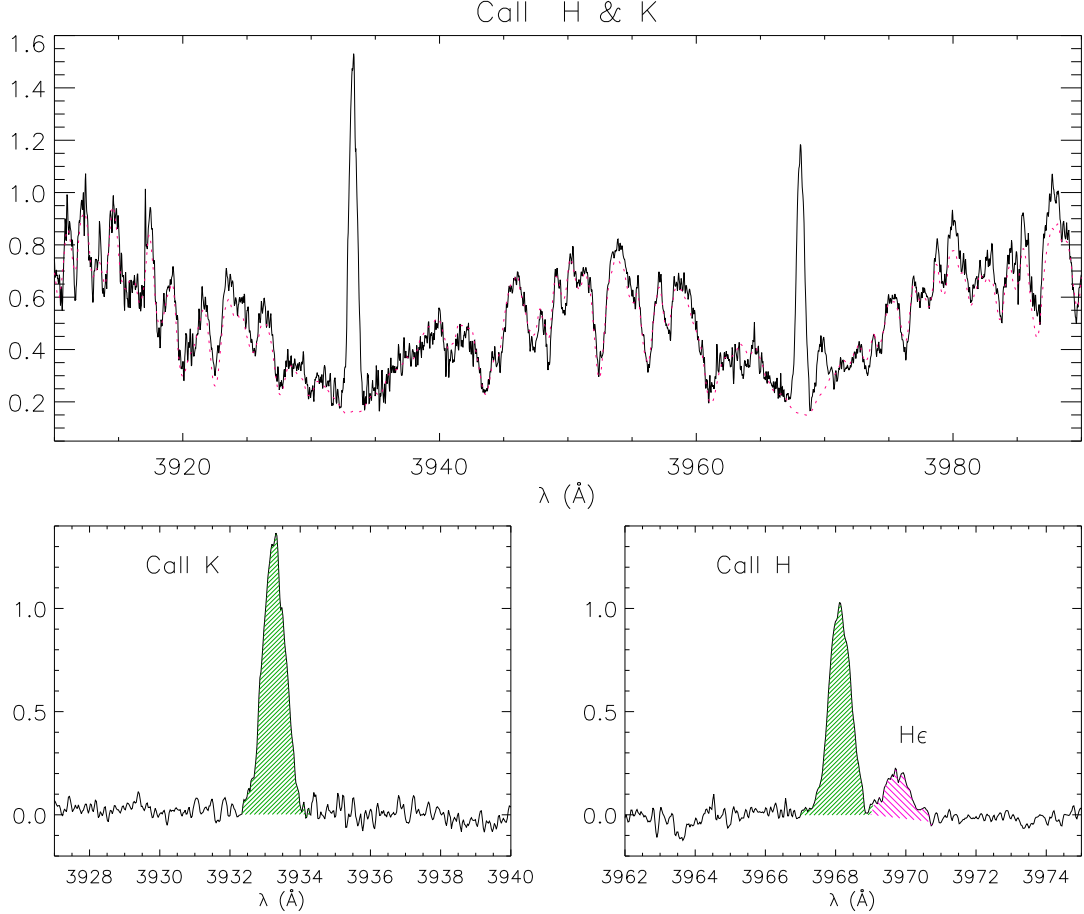} 		
\caption{{\it Top panel}: Continuum-normalized SARG spectrum of \kic (solid line) 
in the Ca\,{\sc ii} H\,\&\,K region together with the inactive stellar template (dotted red line) 
broadened at the $v\sin i$ of the target and Doppler-shifted according to the $RV$ difference. 
{\it Bottom panels}: The difference spectrum. H$\epsilon$ emission, already apparent in the observed spectrum, 
is clearly emphasized in the residuals.} 
\label{fig:CaIIHK}
\end{figure*}

We used the ROTFIT code \citep{Frasca03,Frasca06} to evaluate $T_{\rm eff}$, $\log g$, [Fe/H], and re-determine
$v\sin i$. 
We adopted, as reference spectra, a library of 185 ELODIE Archive spectra of standard stars well 
distributed in effective temperature, spectral type, and gravity, and in a suitable range of metallicities 
(\citealt{Prugniel01}).
The SARG spectrum at $R_{\rm SARG}=57\,000$ taken with the yellow grism was convolved with a Gaussian kernel of width 
$W=\lambda\sqrt{1/R_{\rm ELODIE}^2-1/R_{\rm SARG}^2}$\,\AA\  to match the resolution of ELODIE spectra 
($R_{\rm ELODIE}=42\,000$).
We applied the ROTFIT code to the \'echelle orders with a fairly good S/N, which 
cover the range 4600--6800\,\AA, in the observed spectrum.
For each order, we estimated the stellar parameters of \kic as the mean
of the parameters of the ten reference stars that most closely
resemble the target spectrum, which is quantified by a $\chi^2$ measure,
and adopted their standard deviation as a measure of the
uncertainty.  The adopted estimates for the stellar parameters
come from a weighted mean of the values for all the individual
orders, the weight accounting also for the $\chi^2$ (more weight
to the most closely fitted or higher S/N orders) and the ``amount of information''
contained in each spectral region expressed by the
total spectral-line absorption. The standard error in the weighted
mean was adopted as the uncertainty estimate for the final values
of the stellar parameters.

For the $v\sin i$ determination, we ran ROTFIT using as templates a smaller grid of SARG spectra of inactive
and slowly rotating standard stars (Table\,\ref{Tab:Standards})
acquired with the same instrumental setup as for our target. This allowed us to remove any eventual systematic error 
in the artificial broadening of template spectra introduced by the different resolution and provided us with a
 measure of $v\sin i$ independent of the CCF method. The $v\sin i$ determined with ROTFIT, 
37.1\,$\pm$\,2.9 km\,s$^{-1}$, agrees with the value derived from the CCF.
For these particular tasks, the H$\alpha$ and \ion{Li}{i} lines, as well as spectral regions affected by telluric 
features, were excluded from the analyzed spectral range. 

We also determined the stellar parameters by fitting selected spectral regions with
synthetic spectra generated by SYNTHE \citep{kur81} using ATLAS9 \citep{kur93} atmosphere models.  
When applying all the models, we considered a solar opacity distribution function and microturbulence velocity 
$\xi$\,=\,1~km~s$^{-1}$, which are typical of a star with $T_{\rm eff}$\,=\,5000\,K and $\log g$\,=\,4.5 \citep[see, e.g.,][]{Allende}. 

As starting values of $T_{\rm eff}$ and $\log g$, we used those derived from ROTFIT. At the same time, we determined
the projected rotational velocity by matching the metal lines present in our spectral range.
We adopted the lists of spectral lines and the atomic parameters from \citet{castelli04},
who updated the parameters listed originally by \citet{kur95}.
The stellar parameters were determined by minimizing the difference between the observed and the synthetic
spectra, using the $\chi^2$ as the goodness-of-fit parameter. Errors were estimated to be the variation in the 
parameters that increases the $\chi^2$ of a unit.

In Fig.~\ref{Fig:synthe}, we show the result of our fitting of two different wavelength regions containing important neutral 
iron and calcium lines, which are well suited to constraining temperature, gravity, and metallicity.
The cores of the two strong \ion{Fe}{i} lines at $\lambda$5227.2 and $\lambda$5232.9 are possibly influenced by NLTE
effects and chromospheric filling.  

In Table~\ref{Tab:StarChar}, we list the stellar parameters derived with both ROTFIT and SYNTHE, which agree very well with 
each other.

\subsection{Chromospheric activity and lithium abundance}
\label{Sec:Chrom}

\begin{figure*}[th]
\centering
\includegraphics[width=18cm]{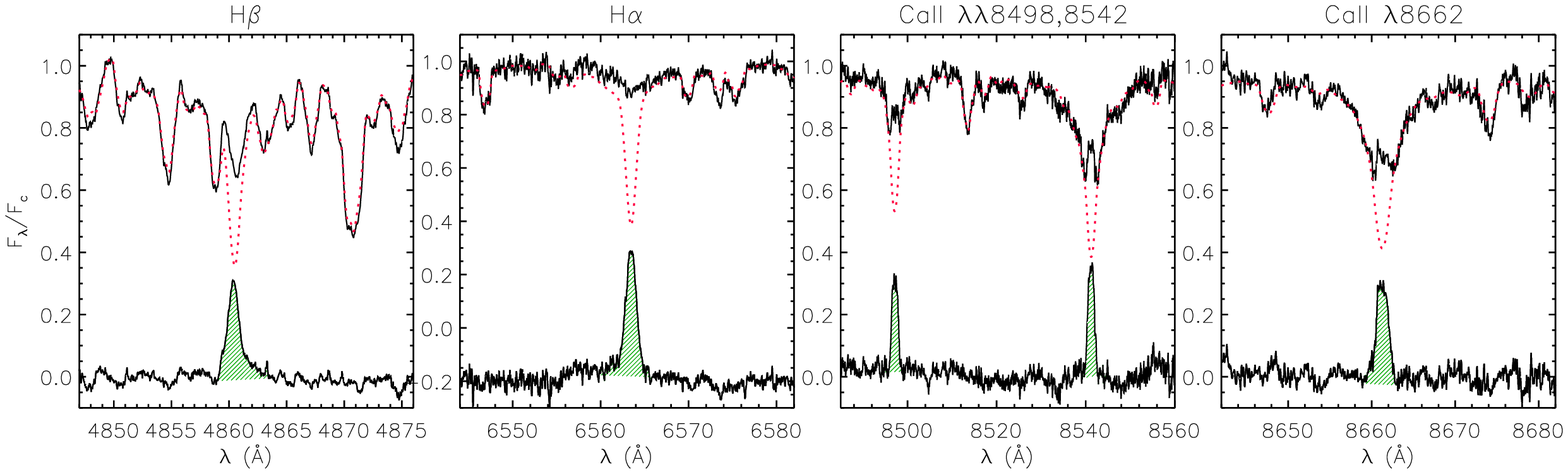}		
\caption{{\it Top of each panel}: Observed, continuum-normalized SARG spectra of \kic 
(solid line) in the H$\beta$, H$\alpha$, and Ca\,{\sc ii} IRT regions together with the non-active 	
stellar template (dotted red line). {\it Bottom of each panel}: Difference between observed and template spectra.
The residual H$\alpha$ profile is plotted shifted downwards by 0.2 for the sake of clarity.
The hatched areas represent the excess emissions that have been integrated to get the net equivalent widths.} 
\label{Fig:Halpha_IRT}
\end{figure*}

We evaluated the level of chromospheric activity from the emission in the cores of the \ion{Ca}{ii} H\&K and  
IRT lines, as well as in the Balmer H$\alpha$, H$\beta$, and H$\epsilon$ lines.
The star's age was inferred from the lithium abundance. 
We estimated both the lithium equivalent width and the chromospheric emission level with the ``spectral 
subtraction'' technique \citep[see, e.g.,][]{Herbig85,Frasca94}. This technique is based on the subtraction of 
a  ``non-active'', ``lithium-poor'' template made with observed spectra of slowly rotating stars 
of the same spectral type as \kic with a negligible level of chromospheric activity and no apparent lithium absorption 
line. The equivalent width of the lithium line ($W_{\rm Li}$) and the net equivalent width of the \ion{Ca}{ii} H\&K,
\ion{Ca}{ii} IRT, H$\alpha$, and H$\beta$ lines ($W^{\rm em}$)
were measured, in the spectrum obtained after subtracting from the target the non-active template, by integrating the 
residual emission (absorption for the lithium line) profile. 

The \ion{Ca}{ii} H\,\&\,K lines display strong emission cores that appear nearly symmetric without any 
evidence of self-absorption. The H$\epsilon$ emission is clearly visible and becomes more evident after the 
subtraction of the non-active template (Fig.~\ref{fig:CaIIHK}). The peak intensity of the  H\,\&\,K lines
is similar to that displayed by \object{SAO~51891}, a very young K0--1 star studied by \citet{Biazzo09}.

In Fig.~\ref{Fig:Halpha_IRT}, the spectral subtraction reveals \ion{Ca}{ii} IRT lines with emission reversals in their
cores and a remarkable filling-in of the H$\beta$ line. Furthermore, the 
H$\alpha$ line is completely filled-in by emission, with high values of net equivalent width (see Table~\ref{tab:fluxes}).
This behavior is typical of very active stars and the H$\alpha$ profile is quite similar to those frequently 
displayed by \object{LQ~Hya}, a well-studied young star for which the H$\alpha$ changes from filled-in 
to weak emission profiles \citep{Fekel86,Strassmeier93,Frasca08}. 
The similarity between KIC~8429280 and LQ~Hya goes beyond the H$\alpha$ profile, since the spectral type is
the same for both stars (K2\,V), the effective temperature is nearly identical, and the rotation period of LQ~Hya 
is only slightly longer (1.6 versus 1.2 days).

We also evaluated the total radiative losses in the chromospheric lines following the guidelines of \citet{Frasca2010}, i.e. 
by multiplying the average $W^{\rm em}$ by the continuum surface flux at the wavelength of the line.	
The latter was evaluated by means of the spectrophotometric atlas 
of \citet{GunnStryker} and the angular diameters calculated by applying the \citet{Barnes1976} relation. 
The net equivalent widths and the chromospheric line fluxes are reported in Table~\ref{tab:fluxes}.

\begin{table}
\caption{Line equivalent widths and radiative chromospheric losses.}
\centering
 \begin{tabular}{lcccc}
  \hline\hline
  \noalign{\smallskip}
  Line                  & Date  & $W^{\rm em}$  &  Error  & Flux                \\  			     
                        & (yyyy/mm/dd)  & (\AA)	  & (\AA)   &(erg\,cm$^{-2}$\,s$^{-1}$) \\
  \noalign{\smallskip}
  \hline
  \noalign{\smallskip}
  H$\alpha$             & 2007/05/31   & 0.922 & 0.107 & 3.35$\times10^6$   \\
    "	                & 2009/07/13   & 1.026 & 0.150 & 3.73$\times10^6$   \\
    "	                & 2009/08/11   & 0.999 & 0.071 & 3.63$\times10^6$   \\
  H$\beta$              & 2009/07/13   & 0.315 & 0.118 & 1.16$\times10^6$	\\
    "                   & 2009/08/11   & 0.442 & 0.076 & 1.63$\times10^6$	    \\
  H$\epsilon$           &  " ~~~"    & 0.264 & 0.040 & 0.48$\times10^6$   \\
  \ion{Ca}{ii} H        & " ~~~"    & 1.075 & 0.068 & 1.66$\times10^6$    \\
  \ion{Ca}{ii} K        & " ~~~"    & 0.795 & 0.045 & 1.45$\times10^6$    \\
  \ion{Ca}{ii} IRT $\lambda8498$   & " ~~~"   & 0.538 & 0.065 &  1.45$\times10^6$   \\
  \ion{Ca}{ii} IRT $\lambda8542$   & " ~~~"   & 0.624 & 0.065 &  1.69$\times10^6$   \\
  \ion{Ca}{ii} IRT $\lambda8662$   & " ~~~"   & 0.596 & 0.074 &  1.63$\times10^6$   \\
  \noalign{\smallskip}
  \hline
\end{tabular}
\label{tab:fluxes}
\end{table}

On the basis of the H$\alpha$ and H$\beta$ flux, we evaluated a Balmer decrement
$F_{\rm H\alpha}/F_{\rm H\beta}$ in the range 2.2--3.2.
Values of the Balmer decrement in the range 1--2 are typical of optically thick emission by solar and stellar
plages \citep[e.g., ][]{Chester91,Buza89}, while prominences seen off-limb give rise to values of  $\sim 10$.
The flux ratio of two \ion{Ca}{ii}-IRT lines, $F_{8542}/F_{8498}=1.2$, indicative of high optical depths, is in the 
range of the values found by \citet{Chester91} in solar plages. 
Solar prominences have instead values of $F_{8542}/F_{8498}\sim 9$, typical of an optically-thin emission source. 

This suggests that the bulk of chromospheric emission of KIC~8429280, in both the Balmer and the \ion{Ca}{ii} lines, is basically 
due to magnetic surface regions similar to solar plages, and eventually prominences will play a marginal role.

\begin{figure}[htp]
\includegraphics[width=8.5cm]{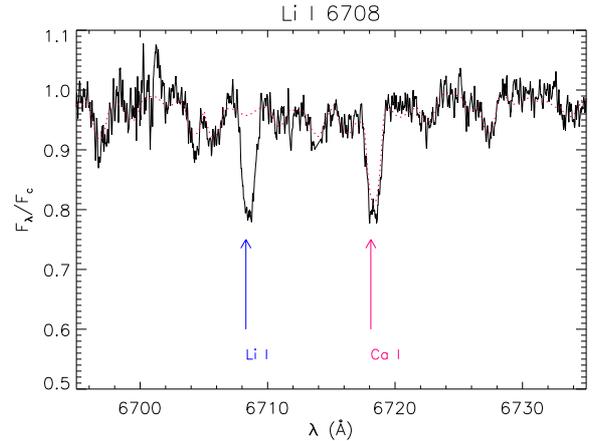}		
\caption{Continuum-normalized spectrum (solid black line) in the lithium  region of KIC~8429280 together 
with the template spectrum (dotted red line) of the non-active lithium-poor reference star, 
broadened at the $v\sin i$ of the target and Doppler-shifted according to the $RV$ difference. 
}
\label{Fig:lithium}
\end{figure}

The subtraction method also permits the measurement of the lithium equivalent width cleaned up from the 
contamination of the close \ion{Fe}{i} $\lambda\,6707.4\,$\AA\  photospheric line (Fig.~\ref{Fig:lithium}). 
The lithium equivalent width, $W_{\rm Li}=270\pm20$\,m\AA, above the Pleiades (100\,Myr) upper envelope 
\citep{Soderblom1993}, translates into a very high lithium abundance, $\log N{\rm(Li)}\simeq2.9\pm0.1$, 
adopting the calibrations proposed by \citet{PavMag96}. With this lithium abundance and $T_{\rm eff}=5055$\,K,
the star falls on the upper envelope of the $\alpha$\,Per cluster (50\,Myr, \citealt{Sestito05}), indicating an age of 
$\approx 50$\,Myr. 

For comparison, \citet{Fekel86} measured for LQ~Hya a slightly lower lithium equivalent width, 
$W_{\rm Li}=234$\,m\AA, deducing a lithium abundance $\log N{\rm(Li)}\simeq2.8$, and concluded that 
that star should be at least as young as the Pleiades.
This supports the measurement of a very young age for KIC~8429280, which could still be in the wTTS or pTTS
phase and not yet have reached the ZAMS.

\subsection{Spectral energy distribution}
\label{Sec:SED}
%

\begin{figure}
\centering{\hspace{-0.4cm}\includegraphics[width=8.5cm]{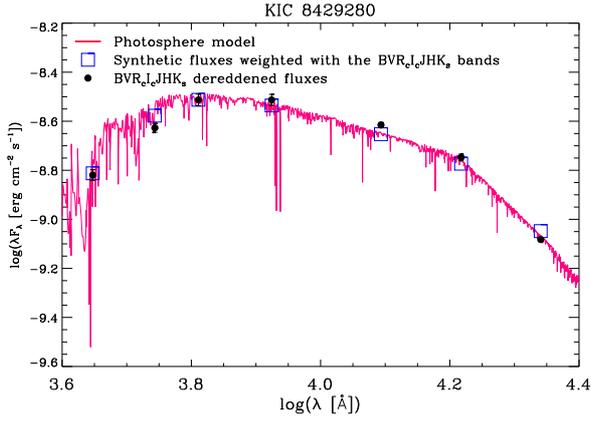}} 		
\caption{Spectral energy distribution for KIC~8429280 (dots). 
The NextGen spectrum with $T_{\rm eff}=5000$\,K and $\log g=4.0$ is overplotted with a continuous line. } 
  \label{Fig:SED}
\end{figure}

The standard $BVR_{\rm C}I_{\rm C}$ photometry (Table~\ref{Tab:StarPhot}), complemented with $JHK_{s}$ magnitudes 
from the 2MASS catalog \citep{2MASS}, allowed us to reconstruct the spectral energy distribution (SED) from the 
optical to the near infrared (IR) domain. 

To perform a fit to the SED, we used the grid of NextGen low-resolution synthetic spectra \citep{Hau99a} with 
$\log g = 4.0$ and 4.5 and solar metallicity. The effective temperature was set to 5000\,K in agreement 
with the value found with ROTFIT and SYNTHE (Table~\ref{Tab:StarChar}).
The interstellar extinction ($A_V$) was evaluated from the distance according to the rate of 0.8 mag/kpc found by 
\citet{Miko} for the sky region around CI~Cyg (very close to our target). 
Finally, the angular diameter ($\phi$), which scales the model surface flux over the stellar flux at Earth, was  
allowed to vary.
The \citet{Cardelli89} extinction law with $R_V=3.1$ was used.
The best-fit solution was found by minimizing the $\chi^2$ of the fit to the $BVR_{\rm C}I_{\rm C}J$ data, which are 
dominated by the photospheric flux of the star and are normally not appreciably affected by infrared excesses. 
The angular diameter derived from the SED, $\phi=0.140$\,mas, implies a distance $d\simeq 50$\,pc if we adopt a 
ZAMS radius of 0.75\,$R_{\sun}$. Adopting a radius of 0.88\,$R_{\sun}$, which is typical of a 50-Myr old star 
with the same temperature as \kic (see Sect.\,\ref{data}), a distance of about 60\,pc is derived.  

As seen in Fig.~\ref{Fig:SED}, the SED is quite well reproduced by the synthetic spectrum 
and no excess is visible at near-IR wavelengths. 

This provides a further check of the effective temperature and assures that this star has by far exceeded the 
T~Tau phase, largely dissipating its accretion disk. 
However, a thinner ``debris disk'', whose presence could be only detected by mid- or far-IR data (lacking for 
KIC~8429280), cannot be excluded.

\section{Kepler light curve and spot modeling}
\label{Sec:Spot_mod}

\subsection{Photometric data}\label{data} 

We analyzed the long-cadence data of quarter 0 (Q0) to quarter 2 (Q2). The 6174 data points are taken from 2009 May 2 to 
2009 September 16, spanning nearly 138 days with a sampling of one photometric point each 30 minutes. 
The most pronounced gap, between Q1 and Q2, amounts to about 4.5 days.

\begin{figure}
\centering{\hspace{-0.4cm}\includegraphics[width=8.5cm]{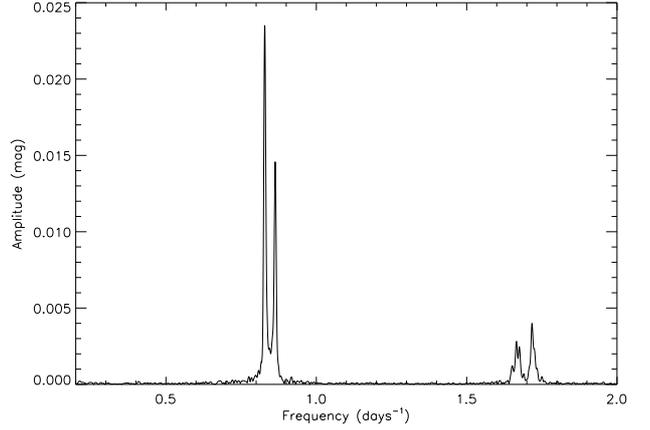}}  	
\caption{Power spectrum of the \kep Q0+Q1+Q2 time series of KIC~8429280. } 
  \label{Fig:power}
\end{figure}

The power spectrum of the \kep time series, displayed in Fig.~\ref{Fig:power}, shows two main peaks very close 
in frequency (0.828 and 0.863 d$^{-1}$), which correspond to periods of about 1.207 and 1.159 days, respectively.
The period uncertainty, estimated from the FWHM of the spectral window, is about 0.006 days.
The two low-amplitude peaks at frequency of $\approx$\,1.66 and 1.72 d$^{-1}$ are overtones of the two main peaks.

The double-peaked periodogram is a clear fingerprint of differential rotation, as shown, e.g., by \citet{Lanza94} in
their simulation of light curves of differentially rotating spotted stars. 
They show that a Fourier analysis of uninterrupted time series of photometry with high precision	
($\frac{\Delta F}{F}=10^{-5}-10^{-6}$, which is comparable with the precision of the \kep data) may detect a 
Sun-like latitudinal differential rotation.

A parameter needed to apply a spot model to our data is the inclination of the rotation axis with respect
to the line of sight. From the values of $v\sin i$, stellar radius $R$, and rotation period $P$ the inclination 
of the rotation axis can be estimated as 
\begin{equation}
\label{Eq:vsini}
\sin i = \frac{v\sin i \cdot P}{2\pi R}\,. 
\end{equation}
{\noindent  
The stellar radius cannot be estimated from effective temperature and luminosity, because this star has no 
Hipparcos parallax and the Tycho value is unreliable. From Eq.~\ref{Eq:vsini}, a lower limit follows, 
$0.85\,R_{\sun}$, which exceeds the  radius of a ZAMS star with $T_{\rm eff}=$\,5055\,K and a mass of 0.76\,$M_{\sun}$, 
$R\simeq 0.75\,R_{\sun}$.}
Thus, we considered a radius of a younger star at the same temperature. We found, from the evolutionary
tracks of \citet{Siess00}, that a $\sim$50\,Myr star of 0.9\,$M_{\sun}$ has the same temperature of
our target, but a radius of 0.88\,$R_{\sun}$. With this value, we find  
$\sin i \simeq 0.965$, which
implies an inclination of about 75$\degr$. 
Such a high inclination is confirmed by the value of $i\simeq\,70\degr$ derived from the light-curve solution 
(see Sect.~\ref{Results:spots}).

\subsection{Bayesian photometric imaging}\label{Bayes} 

Photometric imaging can be a many parameter problem.
An efficient and robust method is light-curve fitting using circular spots. 
We chose to use Dorren's (1987) analytical starspot model in its generalization to 
a quadratic limb-darkening law. 
Any departure from the spherical shape caused by the star's fast rotation was neglected.

In the case of seven spots, we have 46 free parameters. Three parameters describe 
the star: inclination angle $i$, equatorial angular velocity $\Omega_{\rm eq}$, and equator-to-pole 
differential rotation $d\Omega$. To reduce the number of free parameters and enforce a 
monotonic dependence on latitude, $\beta$, differential rotation is 
prescribed by a Sun-like $\sin^2$-law: 
\begin{equation}
\Omega(\beta) = \Omega_{\rm eq} - d\Omega\sin^2\beta\,.
\label{eq:omega}
\end{equation}
{\noindent An additional parameter describes the  spot rest intensity 
$\kappa$. For each spot, six parameters are needed: the longitudes 
in an arbitrarily rotating 
reference system at the beginning and at the end of the observation, 
the spot area at a given time, and three further parameters describing 
spot area evolution (for details see below). The two coefficients $u^+$  
and $u^-$ in the quadratic limb-darkening law \citep[cf.][]{alonso} were 
fixed using the tabulated coefficients from stellar model atmosphere calculations \citep{sing}:    
$u^+ = 0.7042$ and $u^- = 0.3774$. The limb darkening is assumed to be the same 
for the unperturbed photosphere as well as for the spots. }

The parameterized law of differential rotation (Eq.~\ref{eq:omega})
provides the absolute value of a spot's latitude $\beta$ from its rotational 
frequency. To find out the hemisphere to which a spot belongs, we made a test run with 
128 Markov chains in parallel, trying out all possibilities to arrange 
seven spots in the two hemispheres. 
Since the inclination was allowed to take negative values, there are always two
Markov chains describing the same spot allocation situation. From this
64-hour test run, the three different solutions mentioned below
(Sect.\,\ref{Results:spots}) emerged. It cannot be excluded for certain that there are further
solutions.

Spot evolution is described by a simple {\it ansatz\/}: The spot area, 
expressed in units of the star's cross-section, 
is assumed to evolve basically linearly with time. For a decaying spot, this 
prescription allows in principle an estimate of the turbulent 
magnetic diffusivity. The spot evolution model even allows for one sudden 
change in the slope of the area-versus-time relation. Hence, 
altogether four parameters are needed to describe the area of a spot and its 
evolution: area at the time of the (sharp) bend, this time itself and two time 
derivatives of the spot area. 

In addition to the free parameters of the model, the derived parameters, whose physical
meaning is relevant, are considered. We are interested in their marginal distributions, 
in order to derive the mean values and error bars. 
Examples are the latitude and period of a spot. 
The latter simply follows from the initial and final longitude of the spot, the 
former (except for its sign, i.e. the stellar hemisphere on which the spot is located) 
from the spot's rotation period and the two coefficients of the $\sin^2$-law for the differential rotation.

To estimate all these parameters from the data, 
a Bayesian method was applied. This provides 
a straightforward way of estimating parameters and their uncertainties from the data alone. 
Before analyzing the data, proper prior distributions have to be assigned. 
If information about the star's 
inclination $i$ is lacking and selection effects are ignored,  
$\cos i$ should be evenly distributed for  
$0 \le i \le \pi/2$. 
All non-dimensionless parameters such as radii or periods 
are represented by their logarithms. This ensures that the posterior distribution for a 
radius will be consistent with that of an area and likewise the posterior for a
period with that of a frequency, i.\,e. it makes no difference whether
one prefers radii or areas, periods or frequencies. 

The likelihood function is constructed as follows.
Spotted stars that are geometrically similar exhibit the same light
curve, except for an offset in magnitude. 
This offset is considered arbitrary and removed by integration. 
Only the {\em shape\/} of the light curve matters. Hence, 
it is unnecessary to specify the magnitude of the unspotted 
star, which is anyhow unknown.  
It is also assumed that the measurement
errors have a Gaussian distribution.	
The integration over the offset can then be 
done analytically, which is quite convenient.

With the $N$ data points $d_i$ gained at times $t_i$,  
their standard deviation $\sigma$, the fit
$f_0(t_i,p_1\dots{}p_M)$ and an offset $c_0$, the likelihood reads

 $$\Lambda\left(\sigma, c_0, p_1\dots{}p_M; d_i\right) = $$
 $$\prod_{i=1}^{N}{1\over\sqrt{2\pi}\sigma}\exp\left(-{\left(d_i-f_0(t_i,p_1\dots{}p_M)
  - c_0\right)^2\over 2\sigma^2}\right).$$ \\
The $M$ unknown parameters are denoted by $p_j$, with $j=1\dots{}M$. 
Integrating analytically away first the measurement error $\sigma$ -- using
Jeffreys' $1/\sigma$-prior \citep[cf. ][]{Kass96} -- and then the uninteresting offset $c_0$, one gets
a likelihood depending on spot modeling parameters $p_1\dots{}p_M$ only.
It is this {\em mean\/} likelihood, averaged appropriately over measurement
error and offset, from which the posterior density distributions for all unknowns 
are obtained by marginalization. 
We can argue that the method itself determines error and
offset from the data.

When the light curve consists of different parts, first the likelihood of 
each individual part is determined. Afterwards all likelihoods are multiplied. 
This means, each part has been assigned its own error level and offset.

We note that, besides an offset, a linear trend in the data may be taken 
into consideration, where the slope is also integrated away analytically.

The Markov chain Monte Car\-lo (MCMC) method \citep[cf.][]{press}  
has been applied to explore the likelihood mountain in the high-dimensional parameter space.  
 Basically, MCMC is able to find the globally best solution.

\subsection{Results}\label{Results:spots}
 
There are six gaps exceeding a few hours in the combined light curve. 
Accordingly, the light curve has been divided into seven parts, with 
each part being assigned its own offset and error level. Hence, 
artificial vertical jumps between the data sets do not matter.

The MCMC method found three solutions, all with the same level of residuals. 
Two of the three solutions are quickly relaxed. 
The non-relaxed solution needs in any case very dark spots with spot intensity around 0.2 
and would lead to a rather high value of the equator-to-pole differential rotation of 
0.35 rad\,d$^{-1}$.  We, therefore, dismiss that solution. Nevertheless, 
all three solutions show nearly the same spot periods, which means that the maximal
observed rotation shear is remarkably the same: $\simeq 0.21$ rad\,d$^{-1}$. 
What is different are the latitudes of the spots and the spot evolution. 
In what follows, we prefer the solution with the lowest degree of differential 
rotation. All figures refer to that solution if not stated otherwise.  
Table \ref{tab01} provides values of some parameters for both relaxed solutions. 

\begin{figure*}
 \includegraphics[width=18cm]{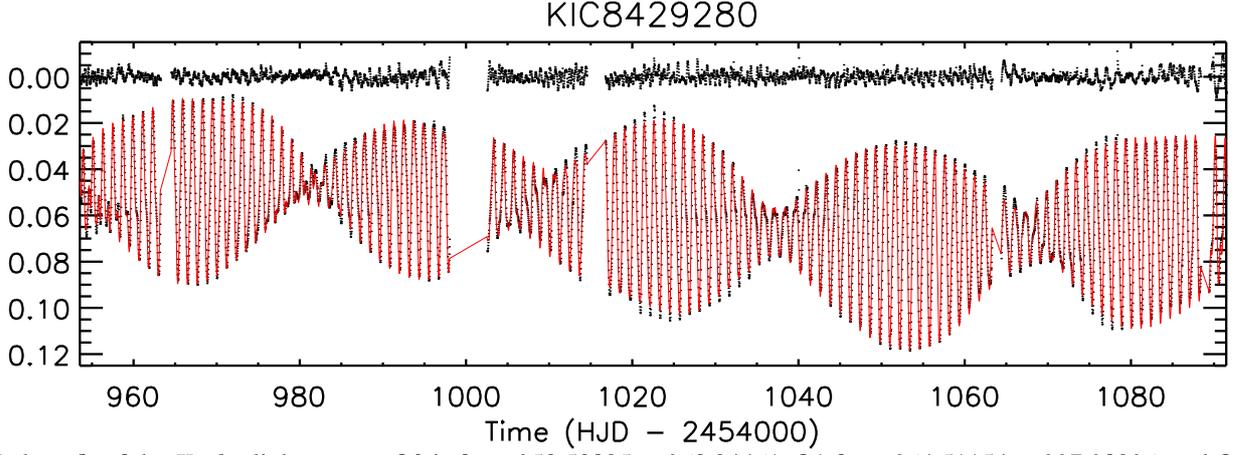}		
\vspace{-70mm}
\caption{ \label{lc}
 Robust fit of the \kep light curves. Q0 is from 
 953.53825 to 963.24461, Q1 from 964.51154 to 997.98296, and 
 Q2 from 1002.76455 to 1091.46678. 
 The solid line is the best fit of the first solution. The residuals, 
 shown at the top, are $\pm 2.36$\,mmag. The gaps are artificially 
 bridged by straight connecting lines.
}
\end{figure*}

\begin{figure}
\begin{center}
 \includegraphics[width=\hsize]{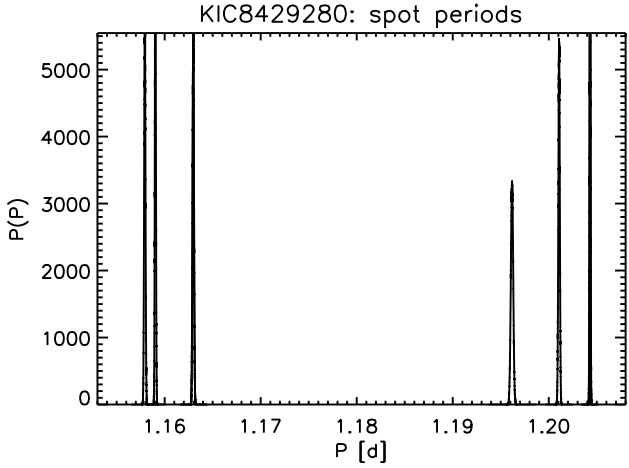}		
\end{center}
\vspace{-5mm}
\caption{ \label{periods}
 Marginal distributions of spot periods.  The two slowest revolving 
 spots exhibit the same period. The periods group around two 
 principal periods, which produces the obvious ``beating'' phenomenon 
 quite understandable. 
}
\end{figure}

 A quick glance at the light curve (Fig.~\ref{lc})  reveals the presence of 
two enduring principal periods. The rotational periods are indeed tightly grouped 
around $P\simeq 1.16$ and $P\simeq 1.20$ days (Fig.~\ref{periods}) 
as already indicated by the power spectrum (Fig.~\ref{Fig:power}).

The stellar surface obviously displays a plethora of long lasting features. Within our 
framework of up to seven {\em dark\/} circular spots, we are unable to fit the light curve down 
to the measurement errors. At the most, we can fit a seven-spot model to the data,  
with the residuals being 2.36 mmag rms. The residuals, exceeding by far the measurement 
errors, are undoubtedly due to the incapability of the model to cope fully  with the data. 

The gain in goodness-of-fit relative to, e.g., a six-spot model is striking. 
To give an example, in a first analysis the combined Q0 and Q1 data 
sets only had been considered. 
In the case of this 45-day subset, the residuals shrink 
from 1.26 mmag to 1.04 mmag if one enlarges the number of spots from six to seven. 
The Bayesian information criterion (BIC) 
clearly justifies the increase in the number of free parameters. 
The BIC \citep{schwarz} allows theory ranking taking into account the 
goodness-of-fit as well as the number of free parameters used. 
With $N$, the number of measurements, $\sigma^2$, the mean variance in one 
measurement, and $M$, the number of free parameters, the criterion reads 
$BIC = N\cdot\log(\sigma^2) + M\cdot\log(N)$. It expresses 
Occam's razor in mathematical terms. A model with a low BIC should preferably be used. 

Attempts to reduce the number of free parameters without any
loss of information (according to BIC) were in vain.  For example, 
we tried to fit the light curve with two different periods only.
In all cases, the loss of goodness-of-fit, as expressed by $\sigma$, 
outclasses the gain in credibility by far.
Thus, one has to accept that -- owing to the unprecedented precision  
of the data -- it is impossible to reduce the number of free
parameters noticeably. That all these free parameters are well 
constrained by the data is testified by the 
mere fact that the Markov chains generally quickly converge to a relaxed solution. 
On the other hand, we did not try to reduce the size of the residuals of the fit by enhancing 
the complexity of the spot model (more spots, different spot temperatures, faculae, etc.), 
because this would require an increase in the number of free parameters with a strong
growth of the computing time and a loss of solution uniqueness. 

\begin{figure}
\begin{center}
 \includegraphics[width=\hsize]{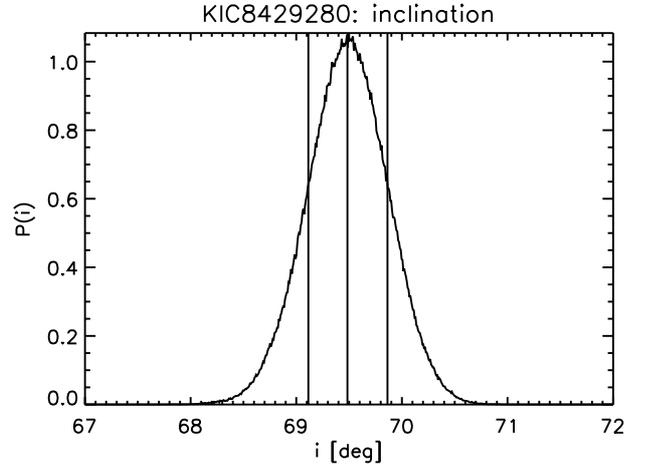}		
\end{center}
\vspace{-5mm}
\caption{ \label{incl}
 The star's inclination from \kep photometry. Vertical lines mark the mean and 
 the 68-per-cent confidence region. The inclination value confirms the 
 relative large radius of the star. 
}
\end{figure}

The marginal distribution for the inclination is shown in Fig.~\ref{incl}.
The inclination angle is photometrically very well constrained, as is
apparent from the smallness of the 68-per-cent confidence region.

\begin{figure}
\begin{center}
 \includegraphics[width=\hsize]{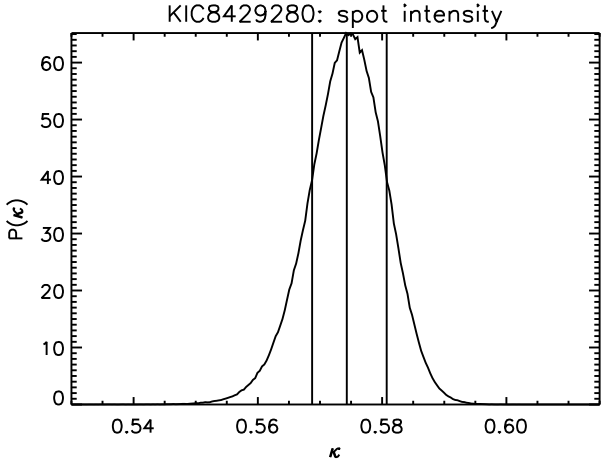}		
\end{center}
\vspace{-5mm}
\caption{ \label{kappa}
 Spot intensity with respect to the unspotted photosphere. 
 Vertical lines mark the mean and the 68-per-cent confidence region.
}
\end{figure}

The spot intensity, measured in units of the unspotted photosphere, 
is $\simeq 0.57$ (cf. Fig.~\ref{kappa}) which sounds reasonable, because,   
in the case of sunspots, the bolometric contrast is 0.67 
\citep[e.g.,][not distinguishing between umbra and penumbra]{Lanza03}. 
For KIC 8429280, a bolometric contrast of 0.57  would translate into a temperature 
difference of $\simeq 650$\,K. 

\begin{figure}
\begin{center}
 \includegraphics[width=\hsize]{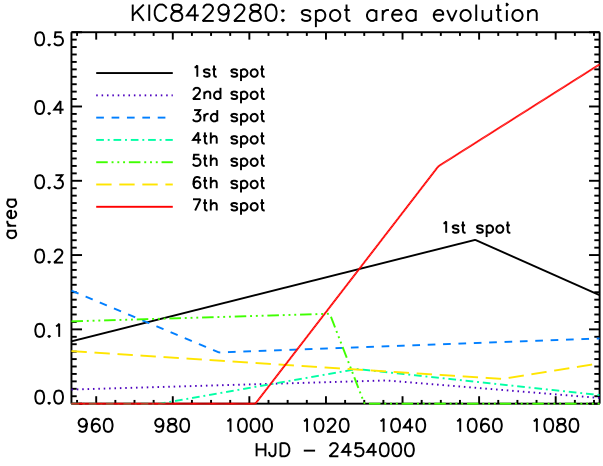}		
\end{center}
\vspace{-5mm}
\caption{ \label{ev304}
 Evolution of spot areas. Areas are expressed 
 in units of the star's cross-section. 
 The largest spots are those at high latitudes.
}
\end{figure}

The spots are rather large and they slowly evolve as appears in Fig.~\ref{ev304}. 
 Despite their large areas, the spots are never really touching each other. 
Hence, spot evolution does not seem to be influenced by the need to avoid spot 
overlapping. One should note that the spot evolution 
{\it ansatz} makes sense within the time span of photometric observation 
only. It must not to be extrapolated to any later times.

\begin{figure}
\begin{center}
 \includegraphics[width=\hsize]{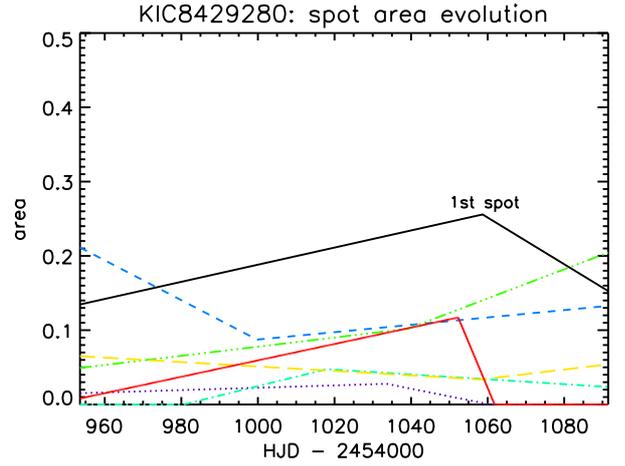}		
\end{center}
\vspace{-5mm}
\caption{ \label{ev306}
 Evolution of spot areas for the second solution. 
 The different types of line have the same meaning as in Fig.~\ref{ev304}. 
}
\end{figure}

As already mentioned, in terms of spot area evolution, the two solutions are 
quite different. We, therefore, provide for the second solution the 
corresponding spot area evolution plot, too (Fig.~\ref{ev306}).

A major difference between the two solutions concerns spots 5 and 7. Spot 7 
of the first solution (the largest one at the end) has partly replaced
spot 5 of the 
second solution. This change in identity is already 
indicated in the periods. 
One should also note that there are, 
apart from spot 3,  small but nevertheless highly 
significant differences in the periods of all  
the other spots (Table\,\ref{tab01}). 
 
That there are at least two solutions is not surprising. 
The inclination is rather high, so it is difficult to 
determine to which hemisphere a spot belongs.

The main result concerns differential rotation.  
In the case of KIC 8429280, the spots are obviously long lasting, 
allowing us to separate the effects of spot evolution from 
differential rotation. 
However, even then a cautionary note is in place 
because of the large number of spots involved. In principle, 
the MCMC method should be able 
to find {\em all\/} possible spot periods, 
in practice one gets at best one feasible set of periods 
-- a conceivable ``scenario'', not more!
The assignment of some dip in the light curve 
to a certain spot is not unambiguous. We lack 
the information about what happens when a spot is out of sight.

\begin{figure}
\begin{center}
 \includegraphics[width=\hsize]{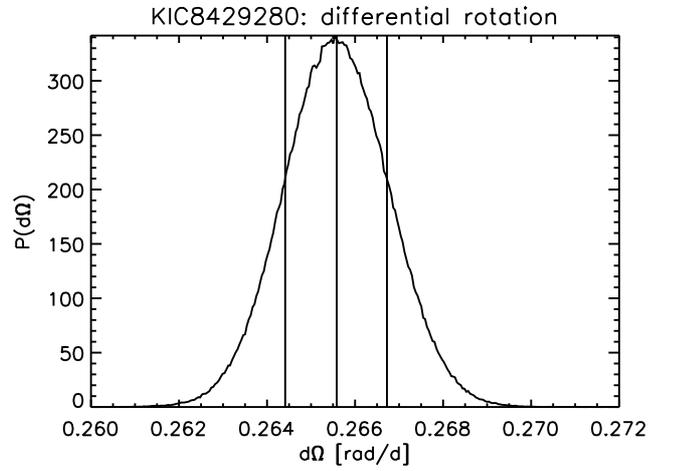}		
\end{center}
\vspace{-5mm}
\caption{ \label{DR}
 The equator-to-pole differential rotation.
 Vertical lines mark the mean and
 the 68-per-cent confidence region.
}
\end{figure}

The marginal distribution of the equator-to-pole differential rotation 
is given in Fig.~\ref{DR}. For our preferred model, it is 
$0.266\pm 0.001$ rad\,d$^{-1}$, given a $\sin^2$ law as a basis 
for the latitudinal dependence of differential surface rotation. 
The second solution results in an even larger 
differential rotation of $0.297^{+0.003}_{-0.002}$ rad\,d$^{-1}$.

As an additional check of the differential rotation, we tried to fit 
the \kep  light curve with a different set of free parameters, i.e. by allowing the 
spot latitudes to vary without any constraint on the differential rotation law.
To save computing time, we applied this model to the Q0+Q1 dataset, which 
is, however, large enough to constrain the rotation period of each spot. 
As a result, we found that  the slowly revolving spots are at high latitudes.
Despite the spot latitudes being notoriously ill-defined by photometry, at
least for ground-based data, the stellar surface rotates beyond any doubt in a Sun-like way.

Expectation values with 1--$\sigma$ confidence limits for
some interesting parameters are given in Table\,\ref{tab01}.
The reader should be aware that the estimated parameter values and their 
(often surprisingly small) error
bars are those of the model constrained by the data. The error bars 
simply indicate the ``elbow room'' of the model, nothing more.

\begin{table}
\centering
\caption{ Two 7-spot solutions: listed are {\em expectation\/} values  
and 1-$\sigma$ uncertainties. Latitudes $\beta$ are derived from 
the assumed law of differential rotation (Eq.~\ref{eq:omega}). Periods $P$ are given in days, 
the spot intensity $\kappa$ is in units of the intensity of the unspotted surface. 
The differential rotation $d\Omega$ (rad\,d$^{-1}$) is the equator-to-pole
value of the shear.}
\label{tab01}

\begin{tabular}{|ll|lc|r|l|r|l|}
\hline
\multicolumn{3}{|c}{parameter} & &\multicolumn{2}{|c|}{1st solution} &\multicolumn{2}{|c|}{2nd solution}\\
\hline
&&&&&&&\\[-8pt]
\multicolumn{2}{|l|}{inclination}&$i$&&$ 69^\circ\hspace{-3pt}.5$ &$^{+0.4}_{-0.4}$         &$ 69^\circ\hspace{-3pt}.6$ &$^{+0.4}_{-0.4}$        \\ [4pt]
1st&    latitude     	&$\beta_1$   &&$ 62^\circ\hspace{-3pt}.5$ &$^{+0.2}_{-0.2}$         &$ 57^\circ\hspace{-3pt}.1$ &$^{+0.5}_{-0.3}$        \\ [4pt]
2nd&    latitude     	&$\beta_2^{\rm a}$   &&$ -0^\circ\hspace{-3pt}.5$ &$^{+0.4}_{-0.2}$         &$ -0^\circ\hspace{-3pt}.8$ &$^{+0.6}_{-0.2}$        \\ [4pt]
3rd&    latitude     	&$\beta_3$   &&$ 62^\circ\hspace{-3pt}.5$ &$^{+0.2}_{-0.2}$         &$ 57^\circ\hspace{-3pt}.5$ &$^{+0.5}_{-0.3}$        \\ [4pt]
4th&	latitude     	&$\beta_4$   &&$ 17^\circ\hspace{-3pt}.3$ &$^{+0.2}_{-0.2}$         &$ 16^\circ\hspace{-3pt}.5$ &$^{+0.2}_{-0.2}$        \\ [4pt]
5th&	latitude     	&$\beta_5$   &&$-53^\circ\hspace{-3pt}.9$ &$^{+0.2}_{-0.2}$         &$ 56^\circ\hspace{-3pt}.7$ &$^{+0.5}_{-0.3}$        \\ [4pt]
6th&	latitude     	&$\beta_6$   &&$ -8^\circ\hspace{-3pt}.0$ &$^{+0.2}_{-0.2}$         &$ 10^\circ\hspace{-3pt}.4$ &$^{+0.3}_{-0.3}$        \\ [4pt]
7th&	latitude     	&$\beta_7$   &&$-58^\circ\hspace{-3pt}.9$ &$^{+0.2}_{-0.2}$         &$-49^\circ\hspace{-3pt}.1$ &$^{+0.2}_{-0.4}$        \\ [4pt]
1st&    period       	&$P_1$    &&$1.20430$		       &$^{+0.00004}_{-0.00004}$ &$1.20390$		     &$^{+0.00004}_{-0.00004}$\\ [4pt]
2nd&    period       	&$P_2$    &&$1.15793$		       &$^{+0.00007}_{-0.00007}$ &$1.15741$		     &$^{+0.00009}_{-0.00009}$\\ [4pt]
3rd&    period       	&$P_3$    &&$1.20430$		       &$^{+0.00005}_{-0.00005}$ &$1.20430$		     &$^{+0.00006}_{-0.00007}$\\ [4pt]
4th&    period       	&$P_4$    &&$1.16298$		       &$^{+0.00007}_{-0.00007}$ &$1.16253$		     &$^{+0.00006}_{-0.00005}$\\ [4pt]
5th&    period       	&$P_5$    &&$1.19617$		      &$^{+0.00012}_{-0.00012}$ &$1.20345$		    &$^{+0.00004}_{-0.00004}$\\ [4pt]
6th&    period       	&$P_6$    &&$1.15903$		      &$^{+0.00005}_{-0.00005}$ &$1.15948$		    &$^{+0.00003}_{-0.00003}$\\ [4pt]
7th&    period       	&$P_7$    &&$1.20107$		       &$^{+0.00008}_{-0.00007}$ &$1.19479$		     &$^{+0.00013}_{-0.00014}$\\ [4pt]
\multicolumn{2}{|l|}{spot intensity}&$\kappa$&&$0.574$            &$^{+0.006  }_{-0.006  }$ &$0.599  $                  &$^{+0.017}  _{-0.015  }$\\ [4pt]
\multicolumn{2}{|l|}{diff. rotation}&$d\Omega$ & &$0.2656$  &$^{+0.0011 }_{-0.0012 }$ &$0.2974 $                  &$^{+0.0016} _{-0.0032}$\\
\noalign{\smallskip}
\hline
\hline 
\end{tabular}
\begin{list}{}{}		         	                	                 
\item[$^{\mathrm{a}}$]  The skewed error estimates for the latitude of the second spot are calculated under the 
assumption that it is located south of the equator.
\end{list}
\end{table}

There is a pronounced long-term trend in the light curve. The star gets dimmer with time. To prevent any instrumental 
effect affecting the outcome of the photometric spot modeling, we allowed not only for offsets, but, 
moreover, for adjustable slopes in the seven time chunks. These slopes are considered nuisance parameters and 
analytically integrated away. The fit is, of course, tighter, with the residuals now being reduced to 2.17 mag rms. 
Most importantly, our main results are unaffected. The inclination is lowered by merely three degrees, the spot 
intensity is slightly increased to 0.635, and the equator-to-pole differential rotation is 
increased somewhat to $0.269\pm 0.003$ rad\,d$^{-1}$, which is between the values of the two
solutions set out in Table\,\ref{tab01}.
Owing to the smallness of these changes, there is no need to be doubtful about the long-term stability of the data. 

An additional improvement could be achieved by including more free parameters in the spot area evolution {\it ansatz}. 
The easiest way to do this is to substitute the sharp bend between the two slopes with a more gradual transition. 
This can be achieved, for example, by a linear interpolation between the values of the two slopes of the 
first derivative of the area-versus-time relation over some time span, which is the new free parameter.
The longer this time, the smoother the bend between the two time evolution regimes.
With seven free parameters more, the residuals are reduced from 2.36 mmag to 2.25 mmag, which is, because of 
the large number of data points involved, formally significant, but the main outcome remains 
nevertheless unchanged. The inclination has decreased by a half of degree to 
$i = 68^\circ\hspace{-3pt}.9\pm 0.4$ and the equator-to-pole differential rotation has decreased to $0.254\pm 0.001$ rad\,d$^{-1}$.

\section{Discussion}
\label{Sec:Disc}

\subsection{Is KIC\,8429280 a binary?}
\label{Sec:Disc:bin}

The constant radial velocity and the absence of secondary peaks in the CCF (Sect.~\ref{Sec:Analysis}) seem to exclude 
that KIC\,8429280 is a spectroscopic binary in a short-period orbit.

However, in the WDS Catalog \citep{Worley97} the source is classified as a close 
visual binary composed of two stars of equal magnitude with a separation of $0\farcs 3$  and a position angle $PA = 22\degr$.
The two components are resolved only in one visual observation performed in 1988 with the 0.5-m Nice refractor 
by \citet{Couteau95}.  
However, although the star is present in the Tycho Double Star Catalog \citep{Fabricius02}, no entry for the 
separation and position angle is present, even though in this catalog binaries with separations $\rho \ge 0\farcs2$ 
and components of similar magnitude are normally resolved. Evidently, the Tycho instrument on-board the Hipparcos 
satellite, which was operational from December 1989 to February 1993 (only 1--5 years later than the Couteau observation),
was unable to distinguish the components of this possible binary. 
As far as we know, no other hint of binarity is present in the literature. 

If we consider two identical stars with the same  luminosity, temperature ($T_{\rm eff}=5055$\,K), and radius ($R=0.88\,R_{\sun}$), 
the distance estimated from the SED increases to about 85\,pc. 
At this distance, an apparent separation of $0\farcs 3$ corresponds to about 25\,AU.
From the third Kepler law, with a mass of about 0.9\,$M_{\sun}$ per each component, we can estimate a minimum orbital period 
$P_{\rm orb}\approx$\,93 years, assuming that the semi-major axis of the orbit is $a\approx a\sin i=25$\,AU. 
Thus, given the long orbital period, the apparent separation of the two components 
should have been nearly the same during the Couteaus's and Tycho observations.   
 
Moreover, in this hypothesis we can evaluate the semi-amplitude of the $RV$ curve as 
\begin{equation}
k = \frac{2\pi}{P_{\rm orb}}\frac{a\sin i}{(1-e^2)^{1/2}} \approx \frac{8}{(1-e^2)^{1/2}}~ \textrm{km\,s$^{-1}$}
\end{equation}

{\noindent were we assume that $a\sin i=25$\,AU.} 
An $RV$ variation should be easily detectable with observations taken a few years apart.

From all the previous considerations, it is unlikely that KIC~8429280 is a double star.
However, more spectra and high angular resolution images are needed to safely exclude the presence of a stellar companion. 

\subsection{Inclination}

The photometrically obtained value of the inclination, $i \simeq 69.5^\circ$, and the 
spectroscopically determined $v\sin i$ allow one to 
determine the radius of the star. Assuming the shortest 
rotational period to be associated with the equator, one arrives at $R = 0.93\,R_\odot$, which is close to 
the expected value (see Sect.\,\ref{Sec:SED}). 
As the inclination depends crucially on the model assumptions, 
this correspondence strengthens our confidence in the 
reliability of our spot-modeling approach. 

\subsection{Spot contrast}

In a first analysis, considering quarters Q0 and Q1 only, 
we identified rather large spots with a very low spot-to-photosphere 
contrast ($\kappa \simeq 0.8$).
After analyzing a three-times longer time span, the spot surface 
brightness (Fig.~\ref{kappa}) in the \kep photometric band
amounts to the comforting value of 60\,\% of the photospheric intensity. 
The corresponding temperature contrast $T_{\rm sp}/T_{\rm ph}$ 
is 0.87.

In the case of 4 PMS stars in Orion ($\sim$\,2 Myr old), \citet{Frasca09} 
found a temperature contrast $T_{\rm sp}/T_{\rm ph}$ in the range 0.71--0.92. For fairly 
young stars ($\sim$\,200--400 Myr) such as $\kappa^1$~Cet, HD~166, and $\epsilon$~Eri, 
\cite{Biazzo07} derived values of $T_{\rm sp}/T_{\rm ph} \sim$\,0.85  from the 
simultaneous modeling of temperature (from line-depth ratio analysis) and light curves.
The spots of  KIC8429280  are, therefore, comparable to both those of Orion PMS and mildly active stars.
However, we recall that a ``spot'' can be more than a single dark spot, namely an active region, which 
is a mixture of dark and bright surface features.

However, despite their very high precision, the \kep magnitudes are taken in
``white light'', i.e. with only one very broad band. Thus, there is no possibility 
of constraining the spot contrast without a contemporaneous multi-band photometric or
spectroscopic monitoring.

\subsection{Differential rotation}

One of the unexpected results of the present work is the degree of differential rotation of KIC\,8429280. 
Assuming a  $\sin^2$ law for the latitudinal dependence, one arrives 
at 0.27--0.30 rad\,d$^{-1}$ for the equator-to-pole difference in angular velocity. 
This is about five times larger than in the case of the Sun.

\citet{Barnes05} showed that in general differential rotation increases with stellar effective temperature.
The highest value reported by them for stars as cool as $\sim$\,5000\,K, $d\Omega\approx 0.2$, is that measured 
by \citet{Donati03} for LQ~Hya in 2000.
\citet{Marsden11} extended this dataset with values of $d\Omega$ (in the range 0.08--0.45 rad\,d$^{-1}$)
measured only for stars more massive than the Sun.

For LQ\,Hya, which has similar stellar parameters and is rotating a little bit slower than our target (cf. 
Sect.\ref{Sec:Chrom}), 
\citet{Kovari04} determined the differential surface rotation from the cross-correlation of several 
Doppler maps. A solar-type differential rotation law, i.e. the equator rotating faster than the poles, with 
$d\Omega=0.022$ rad\,d$^{-1}$ (lap time of $\approx$\,280 days) is found by the aforementioned authors.
\citet{Donati03} measured surface differential rotation in LQ\,Hya and AB\,Dor by means of a new Doppler imaging
technique that includes among its free parameters those of a solar-type differential rotation law.
Their results show that the equatorial angular velocity and the amplitude of the latitudinal shear
of AB\,Dor and LQ\,Hya change on timescales of a few years. This is ascribed to the redistribution of the angular 
momentum inside the stars likely to be caused by the action of a non-linear hydromagnetic
dynamo. In particular, they found for LQ\,Hya a decrease in the shear between years 2000 ($d\Omega\simeq 0.20$ rad\,d$^{-1}$) 
and 2001 (0.01 rad\,d$^{-1}$).

\object{HD~141943}, another very young more-massive star that is still in the PMS phase, has  values of  $d\Omega$
ranging from about 0.23 to 0.44 rad\,d$^{-1}$ in different epochs as derived by \citet{Marsden11}.

Very different values of differential rotation have also been found for \object{HD~171488}, a young ($\sim$50 Myr) Sun.
For this star, a very high solar-type differential rotation $d\Omega\approx 0.4$--0.5 rad\,d$^{-1}$, with
the equator lapping the poles every 12--16 days, was found by both \citet{Marsden06} and \citet{Jeffers08}. 
Much weaker values ($d\Omega\approx 0.04$) were derived instead for the same star by \citet{Jarvi08} and 
\citet{Kovari10}.

A weak differential rotation ($d\Omega$ between 0.017 and 0.056 rad\,d$^{-1}$) was found for
$\epsilon$\,Eri by \citet{Froehlich07} from a Bayesian reanalysis of the MOST light curve \citep{Croll2006, Crolletal2006}.  
This is a mildly active star with the same spectral type as \kic (K2\,V), but with a considerably longer rotational period 
($P_{\rm rot}\simeq 11.2$\,days).

Thus, from these observations, it seems that a higher differential rotation is more frequently encountered
in rapid rotators, although different indications come from previous works, mainly based on ground-based photometry, which seem to
display the opposite behavior, i.e. a differential rotation decreasing with the rotation period 
\citep[e.g.,][]{Messina03}. The precision of the ground-based light curves does not allow to draw firm conclusions
and accurate photometry from space, as well as Doppler imaging, is strongly needed to enlarge the database of differential 
rotation measurements that must be compared with the results of new dynamo models for stars in different evolutionary phases (including
the PMS stage) and with both low and high rotation rates.

The high differential rotation that we found for \kic  disagrees with the predictions of the model of \citet{Kueker11}, who instead 
found rather low values of $d\Omega\approx 0.08$ for an (evolved) solar-mass star rotating with a period as short as 1.3\,days.
However, this specific model was made for a star with a remarkably different mass, effective temperature, and inner structure. 

\citet{Covas05} developed dynamo models capable of reproducing variations in surface differential rotation along the activity cycle, 
although they cannot produce anything as extreme as the episodes of almost 
rigid surface rotation ($d\Omega\approx 0.0$) as reported for LQ Hya. On the other hand, these models
can reproduce variations in the differential rotation of several percent for stars with deep convective envelopes rotating
ten times faster than the Sun.
\citet{Lanza06} found that the large value of the rotation shear ($\sim\,$0.2) observed for LQ\,Hya in the year 2000 implies a dissipated 
power that is comparable to the star luminosity, but this high shear can be maintained for a time of a few years.

The value of 0.27--0.30 rad\,d$^{-1}$  we derive for \kic is the largest ever measured, to our knowledge,
for a young star less massive than the Sun. The analysis of \kep photometry acquired in subsequent epochs will enable 
us to detect any eventual variation in the surface differential rotation.

\section{Conclusion}
\label{Sec:Conc}

We have performed an accurate analysis of high-resolution spectra and high-precision \kep photometry of 
\kic  designed to characterize the chromospheric and photospheric activity. 

Thanks to the high-resolution spectra we have derived, for the first time, the astrophysical 
parameters, $T_{\rm eff}$, $\log g$, [Fe/H], rotational and heliocentric radial velocity, and lithium abundance. 
The lithium abundance allows us to estimate an age of about 50\,Myr, that, with a mass of about 0.9\,$M_{\sun}$, 
is likely to imply a pre-main sequence stage with the star approaching the ZAMS along its radiative track.

A high chromospheric activity level was inferred from different spectral diagnostics (H$\alpha$, H$\beta$, \ion{Ca}{ii} H\&K, 
and \ion{Ca}{ii} IRT lines) by applying the spectral subtraction technique.
Both the Balmer decrement, measured as the ratio of H$\alpha$ to H$\beta$ emission flux, and the flux ratio of two 
\ion{Ca}{ii} IRT lines, which is indicative of optically thick emission similar to that observed from extended facular regions.  

The application of a  robust spot model, based on a Bayesian approach and a Markov chain Monte Carlo method, to the high-precision 
\kep photometry spanning 138 days, has allowed us to map the photospheric spots and study their time evolution.
Seven spots were needed to perform a reasonable fit to the \kep light curve, although, because of the exceptional precision of 
the \kep photometry, it is impossible to reach the noise floor without increasing significantly the degrees of freedom and,
consequently, the non-uniqueness of the solution.
The active regions turn out to be mainly located around three latitude 
belts, i.e. around the star's equator and around $\pm (50{\degr}$--60${\degr})$, with the high-latitude spots rotating slower 
than the low-latitude ones. 

The equator-to-pole differential rotation $d\Omega\simeq 0.27$ rad\,d$^{-1}$ is 
one of the largest values ever measured for a star cooler than the Sun. This is at variance with some recent theoretical models
that also predict a moderate solar-type differential rotation for very rapidly rotating main-sequence stars \citep{Kueker11} but 
 could fit the scenario proposed by other modelers of a higher $d\Omega$, which can change along the activity cycle
\citep{Covas05,Lanza06}.

\begin{acknowledgements}

We are grateful to the anonymous referee for a careful reading of the paper and valuable comments. 
This work has been partly supported by the Italian {\em Ministero dell'Istruzione, Universit\`a e  Ricerca} (MIUR) and 
by the {\em Regione Sicilia} which are gratefully acknowledged. 
J.M.-\.Z acknowlegdes the Polish Ministry grant no. N  N203  405139. 
K.B. acknowledges financial support from the INAF Postdoctoral fellowship.
This research made use of SIMBAD and VIZIER databases, operated at the CDS, Strasbourg, France. 
\end{acknowledgements}

\bibliographystyle{aa}

\end{document}